\documentclass[aps,prd,showpacs,nofootinbib,twocolumn,floatfix,superscriptaddress,preprintnumbers]{revtex4}
\usepackage{amsmath}
\usepackage{amssymb}
\usepackage{epsfig}
\usepackage{graphicx}
\usepackage{stmaryrd}

\usepackage{color}

\def\10{$SO(10)$}
\def\21{SU(2) $\otimes$ U(1) }

\def\422{$SU(4) \otimes SU(2) \otimes SU(2)$}
\def\321{SU(3) $\otimes$ SU(2) $\otimes$ U(1)}

\newcommand {\ignore}[1]{}

\def\lsim{\raise0.3ex\hbox{$\;<$\kern-0.75em\raise-1.1ex\hbox{$\sim\;$}}}
\def\gsim{\raise0.3ex\hbox{$\;>$\kern-0.75em\raise-1.1ex\hbox{$\sim\;$}}}

\newcommand{\AddrAHEP}{%
  AHEP Group, Institut de F\'{\i}sica Corpuscular --
  C.S.I.C./Universitat de Val{\`e}ncia \\
  Edificio Institutos de Paterna, Apt 22085, E--46071 Valencia, Spain}

 \newcommand{\ba}{\begin{array}}
\newcommand{\ea}{\end{array}}
\relax
\def\321{$SU(3)\times SU(2)\times U(1)$}

\begin{document}
\preprint{IFIC/09-17}
\renewcommand{\Huge}{\Large}
\renewcommand{\LARGE}{\Large}
\renewcommand{\Large}{\large}

\title{Is the baryon asymmetry of the Universe related to galactic magnetic fields?}
\author{V. B. Semikoz} \email{semikoz@ific.uv.es} \affiliation{\AddrAHEP}
\affiliation{
IZMIRAN, Troitsk, Moscow region, 142190, Russia}
\author{D.D. Sokoloff} \email{sokoloff@dds.srcc.msu.su}
\affiliation{Department of Physics, Moscow State University, 119999,
Moscow, Russia}
\author{J.~W.~F.~Valle} \email{valle@ific.uv.es}
\affiliation{\AddrAHEP}

\date{\today}
\begin{abstract}

  A tiny hypermagnetic field generated before the electroweak phase
  transition (EWPT) associated to the generation of elementary
  particle masses can polarize the early Universe hot plasma at huge
  redshifts $z \gsim 10^{15}$.  The anomalous violation of the
  right-handed electron current characteristic of the EWPT converts
  the lepton asymmetry into a baryon asymmetry.
  Under reasonable approximations, the magnetic field strength
  inferred by requiring such ``leptogenic'' origin for the observed
  baryon asymmetry of the Universe matches the large-scale
  cosmological magnetic field strengths estimated from current
  astronomical observations.
\end{abstract}


\maketitle

\section{Introduction}

Magnetic fields seem crucial for our understanding the
Universe~\cite{Kulsrud:1999bg,Beck:1995zs}, as they may fill
intracluster and interstellar space, affect the evolution of galaxies
and galaxy clusters, playing an important role for the onset of star
formation and determining the distribution of cosmic rays in the
interstellar medium.
Cosmologists support the view that the early Universe indeed hosted
strong primordial magnetic fields which could survive under some
conditions after recombination and serve as seed fields for galactic
dynamo.
The ultimate origin of such fields could be traced to very early phase
transitions predicted by particle physics.

In contrast, the astronomical community tends to believe that we do
not need seed fields in order to explain the origin of large-scale
galactic magnetic fields whose formation is associated to physical
processes in galaxies, protogalaxies
etc~\cite{beck2005cmf,Arshakian:2008cx}.

Fortunately, upcoming radio telescopes such as the Square Kilometre
Array (SKA)~\cite{johnston-2008} are expected shed light on this issue
and help distinguish the two options.

Here we focus on the cosmology camp, further developing the suggestion
made in Refs.~\cite{Semikoz:2003qt,Semikoz:2007ti} where further
alternative realizations are also mentioned. The basic assumption is
the existence of a primordial seed hypermagnetic field, and its
interaction with neutrinos.

Thanks to their unique properties, neutrinos provide the only
``messenger'' capable of probing the early Universe at high redshifts,
$z>z_{\rm recomb}\sim 1100$. The fact that they are required to be
massive in order to account for neutrino
oscillations~\cite{Maltoni:2004ei} opens ways for them to play a role
in cosmology.
For example, if neutrino masses arise via the seesaw
mechanism~\cite{Nunokawa:2007qh} the baryon asymmetry of the Universe
may be easily accounted for through the so-called leptogenesis
mechanism~\cite{fukugita:1986hr}.
Alternatively, the generation of neutrino masses may also shed light
on the dark matter
problem~\cite{Berezinsky:1993fm,Lattanzi:2007ux,Bazzocchi:2008fh}.

Here we consider a tiny primordial seed hypermagnetic field $ B_0^Y$
generated at $T_0\gg T_{EWPT}$. We show that its presence induces a
nonzero lepton asymmetry in the early phases of the evolution of the
Universe which, thanks to the
anomaly~\cite{PhysRevD.14.3432,kuzmin:1985mm}, can be ``leptogenic''
without directly invoking nonzero neutrino masses.
At present times such asymmetry can exist only in neutrinos and its
possible detection remains a challenge~\cite{Gelmini:2004hg}.
However it may have cosmological implications for the cosmic microwave
background~\cite{Lesgourgues:1999wu} as well as for Big Bang
nucleosynthesis. For example the latter constrain the electron
neutrino asymmetry, $\mid \xi_{\nu_e}(T_{BBN})\mid\lsim 0.07$,
$T_{BBN}\sim 0.1~{\rm MeV}$. Thanks to the large mixing angles
indicated by neutrino oscillation data~\cite{Maltoni:2004ei} a similar
restriction also applies to the chemical potential of the other
flavors at temperatures $T\sim 3~{\rm MeV}$ \cite{dolgov:2002ab}.

It is especially interesting to consider the possible effects of a
neutrino asymmetry at much earlier times, say, at $T\gg 3~{\rm MeV}$
or $z \gsim 10^9$. It is easy to generalize Maxwell's equations for
the hypercharge field $Y_{\mu}$ present in the Standard Model by the
addition of the parity violation pseudovector current term ${\bf
  J}_5\sim \alpha {\bf B}_Y$, where
\begin{equation}\label{eq:munu} \alpha
(T)=\frac{47g^{'2}\mu_{\nu}(T)}{1512 \pi^2 \sigma_{\rm cond}(T)},
\end{equation}
$\mu_{\nu}(T)=\sum_{l=e,\mu,\tau}\mu_{\nu_L}^{(l)}(T)$ being the
net neutrino chemical potential, and $\sigma_{\rm cond}(T) \sim T$
denotes the hot plasma conductivity.
A nonzero net neutrino asymmetry $\xi_{\nu}^{(l)}$ results, which may
lead to a strong amplification of the primordial hypermagnetic field
in the early Universe hot plasma~\cite{Semikoz:2007ti}, e.g.
\begin{equation}
\label{eq:growth}
B_Y(x) = B_0^Y\exp\left[32\int_x^{x_0}\frac{{\rm d}x^{'}}{x^{'2}}\left(\frac{\xi_{\nu}(x^{'})}{0.001}\right)^2\right]
\end{equation}
where we introduced the new variable $x=T/T_{EWPT}$ and $B_0^Y$ is the
assumed amplitude of the seed hypermagnetic field. Note that this
result relies only on the standard plasma physics and Standard
particle physics, as it follows simply from the basic parity violating
nature of the Standard Model.

Neglecting in MHD the displacement current $\partial {\bf
  E}_Y/\partial t$ and using Maxwell's equation $\partial_t{\bf
  B}=-\nabla \times{\bf E}_Y$ we easily derive Faraday's equation in
the rest frame of the Universe ${\bf V}=0$, as
 \begin{equation}\label{eq:Faraday}
\partial_t{\bf B}_Y=\nabla\times \alpha {\bf B}_Y + \eta \nabla^2{\bf B}_Y,
\end{equation}
where $\eta =(4\pi \sigma_{\rm cond})^{-1}$ is the hypermagnetic
diffusion coefficient.

We now turn to discuss the subsequent evolution of the asymmetry. As
the theory passes from the unbroken gauge symmetry phase to the broken
one, the anomalous violation of fermion number plays a key
role~\cite{PhysRevD.14.3432,kuzmin:1985mm}.
In the presence of the anomaly for right-handed electrons one
has~\cite{Giovannini:1997eg}
 \begin{equation}
 \label{eq:Abelian}
\partial_{\mu}j^{\mu}_{eR}=-\frac{g^{'2}y_R^2}{64\pi^2}Y_{\mu\nu}\tilde{Y}^{\mu\nu},~~~y_R=-2,
\end{equation}
so that~\cite{PhysRevD.49.6394},
 \begin{eqnarray}
 \label{eq:system}
&\frac{{\rm d}L_{e_R}}{{\rm dt}} \approx
-\frac{g^{'2}}{4\pi^2 s}{\bf
E}_Y\cdot{\bf B}_Y\,, \quad  2\frac{{\rm d}L_{e_L}}{{\rm
dt}}=-\frac{{\rm d}L_{e_R}}{{\rm dt}} +\frac{1}{3}\dot{B}
\end{eqnarray}
where $s=2\pi^2g^*T^3/45$ is the entropy density; $L_l=(n_l -
n_{\bar{l}})/s$, $B=(n_B-n_{\bar{B}})/s$ are the lepton and baryon
numbers correspondingly and we have neglected the collision integrals
associated with decay (inverse decay) of Higgs bosons,
e.g. $\phi^{(0)}\leftrightarrow e_L\bar{e}_R$.

Substituting the conservation law $L_{eR}=B/3 - 2L_{eL}$ that follows
from the second equation in (\ref{eq:system}) into the first equation
in (\ref{eq:system}), and taking into account the adiabatic
approximation, $s\approx {\rm const}$, so that chemical potentials
change very slowly, $
\partial_t \mu_{eL}=\partial_t \mu_{\nu_{eL}}\approx 0,~~{\rm or}~~{\rm d}L_{eL}/{\rm dt}\approx 0,
$
one gets the change of the baryon asymmetry in the presence of
hypercharge fields as,
 \begin{equation}
 \label{eq:baryonchange}
\frac{1}{3}\frac{\partial (n_B - n_{\bar{B}})/s}{\partial t}= - \frac{g^{'2}}{4\pi^2s}{\bf E}_Y\cdot{\bf B}_Y.
\end{equation}
Substituting the hyperelectric field ${\bf E}_Y$ from Maxwell's
equations we get the baryon asymmetry at $T_{EW}$ expressed as
 \begin{equation}
 \label{eq:main}
\eta_B (t_{EW})=\frac{3 g^{'2}}{4\pi^2 s}\int_{t_0}^{t_{EW}}\left[\alpha B_Y^2-\eta~(\nabla\times{\bf B}_Y)\cdot{\bf B}_Y \right]{\rm dt},
\end{equation}
where the baryon asymmetry $\eta_B=(n_B-n_{\bar{B}})/s$. This is our
main result.
It also follows by considering the change of the Chern-Simons number
density released in the form of fermions due to the anomaly,
$\eta_B(t_{EW})=(3/2s)\Delta n_{CS}$, where $\Delta n_{CS}$ is given
by
$$\Delta n_{CS}=
-\frac{g^{'2}}{2\pi^2}\int\limits_{t_0}^{t_{EW}}({\bf E}_Y\cdot{\bf B}_Y){\rm dt}.$$

One notes from Eq. (\ref{eq:main}) in order to account for the
observed baryon asymmetry $\eta_B\sim 10^{-10}>0$ one requires a
positive sign for the net neutrino asymmetry,
$\mu_{\nu}=\sum_l\mu_{\nu L}^{(l)}>0$. Note also that the second
diffusion term in Eq. (\ref{eq:main}) must be less than the first one
in $\alpha^2$ dynamo mechanism \cite{1983flma....3.....Z}.

Let us give estimates of baryon asymmetry (\ref{eq:main}) for the
topologically non-trivial hypermagnetic field configuration
\cite{Giovannini:1997eg}, $Y_0=Y_z=0$, $Y_x=Y(t)\sin k_0z$,
$Y_y=Y(t)\cos k_0z$, which leads to exponential amplification of the
amplitude $Y(t)$ (compare with \cite{Semikoz:2003qt,Semikoz:2007ti}),
 \begin{equation}
 \label{eq:exponent}
 Y(t)=Y^{(0)}\exp[\int_{t_0}^{t}[k_0\alpha
(t^{'}) -k_0^2\eta (t^{'})]{\rm dt^{'}}].
\end{equation}
Following Ref.~\cite{Giovannini:1997eg} we find ${\bf
  B}_Y=\nabla\times {\bf Y}=B_Y(t)(\sin k_0z,\cos k_0z,0)$, where
$B_Y(t)=k_0Y(t)$, or we should substitute in Eq. (\ref{eq:main}) ${\bf
  B}_Y^2=B^2_Y(t)$, $(\nabla\times{\bf B}_Y)\cdot{\bf
  B}_Y=B^2_Y(t)k_0$.

Substituting the helicity parameter $\alpha (T)$ given by
Eq.~(\ref{eq:munu}) and keeping all parameters including conductivity
$\sigma_{\rm cond}(T)$ and the hypermagnetic field strength
$B(t_{EW})$ as constants at $T_{EW}$ due to {\it the adiabatic regime
  with entropy $s\approx {\rm const}$ or $T\sim T_{EW}\approx {\rm
    const}$} we estimate the integral in Eq. (\ref{eq:main}) as the
integrand$\times t_{EW}$ where $t_{EW}=(2H)^{-1}=M_0/2T_{EW}^2$,
$M_0=M_{Pl}/1.66\sqrt{g^*}$, or we find from Eq. (\ref{eq:main}):
 \begin{eqnarray}
\label{eq:main1}
&&\eta_B(t_{EW})=\frac{135\alpha^{'}(M_0/2T_{EW})}{
8\pi^4(\sigma_{\rm cond}/T_{EW})g^*}
\left(\frac{B^2_Y(t_{EW})}{T_{EW}^4}\right)\times\nonumber\\&&\times\left[
\left(\frac{ \mu_{\nu}}
{T_{EW}}\right)\frac{47\alpha^{'}}{94.5}-\frac{k_0}{T_{EW}} \right].
\end{eqnarray}
Substituting numbers $\eta_B(t_{EW})\sim 10^{-10}$, $\sigma_{\rm
cond}/T_{EW}\sim 10^2$, $g^*\sim 10^2$, $\alpha^{'}\sim 10^{-2}$,
$M_0/T_{EW}=7\times 10^{15}$, $47\alpha^{'}/94.5\sim 5\times
10^{-3}$ and neglecting the second negative diffusion term one gets
from Eq.~(\ref{eq:main1}) 
 \begin{equation}
 \label{eq:product}
\left(\frac{\xi_{\nu}(T_{EW})}{0.001}\right)\frac{B_Y^2(T_{EW})}{T_{EW}^4}\approx 3.3\times 10^{-14}.
\end{equation}
which constrains the product of hypermagnetic field and asymmetry at
the EWPT, and hence magnitude of the subsequent Maxwellian magnetic
field, obtained from the boundary condition
$A^{(em)}_j=\cos\theta_WY_j$ at EWPT.

In Ref.~\cite{Semikoz:2007ti} a stringent upper bound on the net
neutrino chemical potential $$\xi_{\nu}(T_{EW})/0.001< 0.12$$ was
obtained by requiring field survival against Ohmic dissipation.
This implies a lower bound on the strength of the hypermagnetic field
at EWPT, $B_Y(t_{EW})\gsim 5.24\times 10^{17}~{\rm G}\ll T_{EW}^2\sim
10^{24}~{\rm G}$.  
We use this bound in order to fix the magnitude of
the initial value of the Maxwellian magnetic field $B(t_{EW})\sim
5\times 10^{17}~G$.  

\begin{figure}[b]
\begin{center}
\includegraphics[angle=0,height=7cm,width=0.48\textwidth]{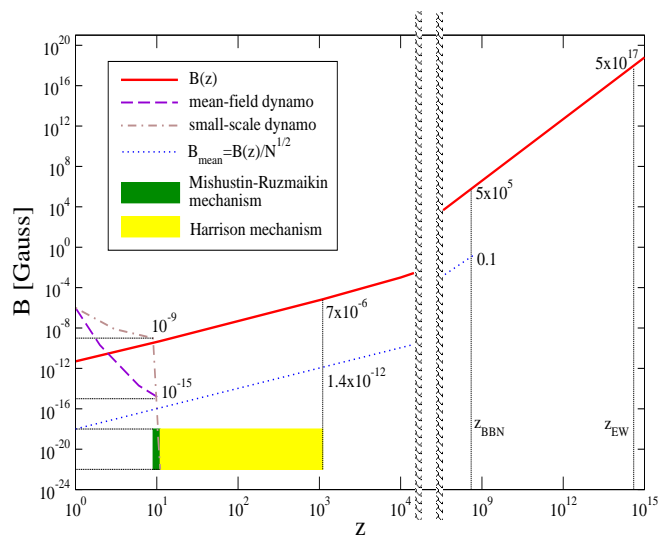}
\caption{Magnetic field evolution after EWPT. The solid (red) line
  represents the Maxwellian magnetic field evolved from the
  hypermagnetic one as frozen-in plasma, while the dotted (blue) line
  represents the large-scale (1 pc at the epoch of galaxy formation)
  component of magnetic field which becomes causal at a moment after
  the EWPT.  The dashed line denotes the galactic magnetic field
  generated by mean-field dynamo, while the dash-dotted one represents
  the galactic magnetic field generated by small-scale and then
  mean-field dynamo, starting from $10^{-9}~{\rm G}$.  For comparison
  we also show in the boxes the magnetic field models in
  Refs.~\cite{boxes}.}
\label{fig:profiles}
\end{center}
\end{figure}

The subsequent evolution of the Maxwellian magnetic field after EWPT
as a result of cosmological expansion is illustrated by the solid
(red) lines in Fig.~\ref{fig:profiles}, in terms of its dependence on
the redshift $z$. The astronomical relevance of this field depends on
its spatial scale. The plot presented can be directly used in order to
estimate the magnitude of the seed field for the galactic dynamo,
provided the field is homogeneous on scales larger than the horizon
size at the epoch of the phase transition.
  Moreover one must assume that the field has been created by a
  noncausal mechanism.

  An alternative option considered here is that the field becomes
  causal at the instant of the phase transition. This means that its
  spatial scale $l$ is very small at later epochs in comparison with
  galactic scales $L_{gal}$ and should be considered as a small-scale
  magnetic field in the context of galactic dynamos. Such magnetic field
  corresponds to smaller large-scale fields,
$$B_{mean}=B N^{-1/2}$$ inferred as statistical mean field,
where $N= (L_{gal}/l)^3$. This magnitude is indicated by the dotted
lines in Fig.~\ref{fig:profiles}.
Note that the mean field $B_{mean}$ is causal at the BBN time,
but not at earlier times, in contrast to the initial Maxwellian field.
This is the reason why the blue (dotted) line for the mean field does
not extend to higher redshits $z>z_{BBN}$.

Both estimates should be compared with the results predicted by the
galactic dynamo theory~\cite{Arshakian:2008cx}. The latter assumes
that first seed fields for galactic dynamos were created at the epoch
of protogalaxies, as indicated by the dashed and dot-dashed lines in
Fig.~\ref{fig:profiles}. One concludes from the avove considerations
that cosmological magnetic fields can at least provide an important,
if not the leading, contribution to the early stages of galactic
magnetic field formation.  \vskip 0.3cm

In summary, here we have considered the effect of a tiny hypermagnetic
field generated by early Universe processes taking place before the
electroweak phase transition. They can polarize the early Universe hot
plasma so that, as the Universe undergoes the EWPT the anomalous
violation of the right-handed electron current converts the lepton
asymmetry into the observed baryon asymmetry.
Under simplifying model assumptions we have inferred the magnetic
field strength at the EWPT by requiring that it reproduces the
observed baryon asymmetry of the Universe. Within this picture one can
also account for the large-scale cosmological magnetic field strengths
estimated from current astronomical observations.

The topologically nontrivial solution Eq.~(\ref{eq:exponent}) can be
reconciled with homogeneity and isotropy of the Universe by
considering a domain structure with topologically nontrivial $\bf
Y$-fields in each domain and random isotropic orientations of the
$z$-axis.  Such Universe is homogeneous and isotropic on scales $L \ll
l_H$ where $L$ is the typical domain size, $l_H=(2H)^{-1}$ is the
horizon size.  Here we have focused our attention on the time
evolution of the hypermagnetic field inside a given domain, the full
picture will be given elsewhere.


\section*{Acknowledgments}

The authors wish to thank Sergio Pastor and Timur Rashba for helpful
discussions.  This work was supported by the Spanish grant
FPA2008-00319/FPA and the CSIC-RAS exchange agreement.  DS
acknowledges financial support from RFBR under grant 07-02-00127a.



\end{document}